\documentclass[aps,prl,reprint,superscriptaddress]{revtex4-2}
\usepackage{graphicx}
\usepackage{amsmath}
\usepackage{amssymb}
\usepackage{bm}

\begin{document}

\title{Kinetic Arrest of a First Order Phase Transition}

\author{Sindhunil Barman Roy}
\affiliation{Department of Physics, Ramakrishna Vivekananda Educational and Research Institute, Belur Math, Howrah 711202 }
\affiliation{Department of Physics, Indian Institute of Technology Bombay, Powai, Mumbai 400076}
\affiliation{UGC-DAE Consortium for Scientific Research, University Campus, Khandwa Road, Indore 452001}


\date{\today}

\begin{abstract}
We report a phenomenological theory for the kinetic arrest (KA) of a first-order phase transition, taking the Mott metal-insulator transition in $V_2O_3$ as a test case. By defining a order parameter $\phi$ related to the monoclinic distortion of the high temperature metallic and mapping its Time-Dependent Ginzburg-Landau (TDGL) dynamics onto a disorder-influenced Imry-Wortis landscape, we derive a universal transcendental condition for the mechanism of the kinetic arrest. We demonstrate that epitaxial substrate-induced clamping in (001)-oriented $V_2O_3$ thin films elevates the elastic activation barriers, trapping the high-symmetry corundum phase down to 4.2~K. This structural suppression of the insulating state robustly explains the observed hysteretic $V$-$I$ switching a hallmark of memristive behaviour. Our work identifies a "Mott-Glass" as a structurally arrested non-equilibrium state in the strained thin-film of V$_2$O$_3$. Our work provides a predictive framework for engineering strain-tuned neuromorphic synapses.
\end{abstract}

\maketitle

First-order phase transitions (FOPTs) are a common occurrence in nature, characterized by discontinuous changes in entropy and either volume, magnetization or polarization as a control variable like temperature, pressure, magnetic field or electric field is varied. While classic examples include the boiling and freezing of water, these transitions are equally significant within solid materials where they involve changes in lattice or spin degrees of freedom. In magnetic solids, such as doped CeFe$_2$ alloys, manganites, and Heusler alloys, these transitions often link magnetic and structural changes, giving rise to multifunctional properties like giant magnetoresistance, the magnetocaloric effect, and magnetic shape memory. Similar transitions are observed in dielectric materials, where they are identified by sharp rises in polarization and field hysteresis, and in the vortex matter of type-II superconductors, where they manifest as melting or solid-solid transitions. A defining feature of FOPTs in these diverse materials is the influence of quenched disorder \cite{imry1979}, which can lead to a landscape of transition temperatures, phase-coexistence, and thermal/magnetic field hysteresis of the thermodynamic variables (see \cite{sbroy2013} and references therein). 

The archetypal Mott insulator $V_2O_3$ exhibits a first-order metal-insulator transition (MIT) at $T_N \approx 160$ K \cite{yethiraj,leiner}. Arising from a delicate competition between electron kinetic energy and Coulomb repulsion \cite{rozenberg,stewart}, this transition is increasingly understood through real-space phase inhomogeneities and nanotextured coexistence of metallic and insulating domains \cite{stewart,lupi,frandsen,macleod}. Such sensitivity to external stimuli positions $V_2O_3$ as a prime candidate for resistive RAM and neuromorphic computing \cite{kalch}. Crucially, the electrically induced insulator-to-metal transition (IMT) offers a pathway to carrier densities far exceeding conventional semiconductors, making the field-driven collapse of the Mott gap a central focus for next-generation electronic devices.

Recent experimental work on epitaxial $V_2O_3$ thin films has revealed interesting electrothermal history effects and tunable memristive behavior, which can be rationalized through a generalized framework of disorder-influenced FOPT combined with resistor network models \cite{kalch2020,de,stolier2014}.The complex phase coexistence observed in $V_2O_3$ and related compounds can be understood as the manifestation of a disorder-broadened FOPT being interrupted by the glass-like arrest of kinetics. This "kinetic arrest" (KA) occurs when the viscous retardation of a transition prevents its completion, resulting in a non-equilibrium state where two phases (such as the high-temperature metallic and low-temperature insulating phases) coexist down to the lowest temperatures \cite{roychaddah}. 

In this letter we present a generalized phenomenological theory of disordered broadened first order phase transition starting from the Ginzberg-Landau model of first order phase transition. To describe the kinetics of the insulator-to-metal transition (IMT), we define a non-equilibrium scalar order parameter, $\phi$, which represents the local degree of metallic phase fraction. Recognizing that the Mott transition in $V_2O_3$ is driven by a structural shift from a monoclinic insulating state to a rhombohedral metallic state, we define $\phi = 1 - \theta$, where $\theta$ (0 $\le \theta \le$ 1) is the structural order parameter representing the monoclinic distortion. In this framework, the purely insulating phase corresponds to $\phi = 0$ ($\theta = 1$), while the high-symmetry metallic phase is represented by $\phi = 1$ ($\theta = 0$). This transformation allows us to directly model the electric-field-induced collapse of the Mott gap while maintaining the underlying structural physics of lattice strain and elastic clamping. We could then reproduce qualitatively all the experimental features including the memristive behaviour arising out of the electric field driven first order Mott insulator to metal transition.   

To describe the complex switching and stalling behavior in $V_2O_3$, we adopt a generalized Ginzburg-Landau (GL) framework. While the Landau theory of first-order phase transitions (FOPT) successfully captures the bulk thermodynamic jump in the order parameter, it remains inherently local and static. In $V_2O_3$, however, the transition is characterized by nanotextured phase coexistence and long-range elastic strain. The GL approach allows us to incorporate these spatial gradients and the influence of quenched disorder, providing a continuous field description of the interface between metallic and insulating regions. Furthermore, by extending this to a Time-Dependent Ginzburg-Landau (TDGL) formulation, we can directly model the kinetics of the phase boundary and the subsequent phase coexistence that leads to the observed memristive history-dependence.

We define a non-equilibrium free energy functional, $F[\phi]$, where $\phi(\mathbf{r}, t)$ is a coarse-grained scalar order parameter. In the context of the Mott transition in $V_2O_3$, $\phi$ represents the metallic fraction ($0 \le \phi \le 1$).The total free energy is expressed as:
\begin{equation}
F[\phi] = \int d^3r \left[ f_{loc}(\phi) + \frac{1}{2}\gamma |\nabla \phi|^2 + V_{dis}(\mathbf{r})\phi \right]
\end{equation}

For a FOPT, the local energy density must exhibit multiple minima. We use a standard polynomial expansion (typically 6th order) to ensure the coexistence of two stable/metastable phases:
\begin{equation}
f_{loc}(\phi) = \frac{A}{2}(T-T_0)\phi^2 - \frac{B}{4}\phi^4 + \frac{C}{6}\phi^6
\end{equation}

Here, $A, B, C > 0$ are the coefficients that define the energy barriers, and $T_0$ is the stability limit of the high-temperature phase. The gradient term ($\frac{1}{2}\gamma |\nabla \phi|^2$) represents the energy cost of the interface between the metallic and insulating phases, and$\gamma$ is the surface tension coefficient, which is crucial for determining the nucleation size of the metallic droplets within the insulating matrix.

To account for the smearing of the FOPT observed in $V_2O_3$, we must move beyond a single, global transition temperature. In real materials, quenched disorder-arising from vacancies, dopant fluctuations, or local dislocations-creates a spatially varying energy landscape. We incorporate the disorder landscape ($V_{dis}$) by treating the local transition temperature, $T_c(\mathbf{r})$, as a stochastic field. Following the Imry-Wortis logic \cite{imry1979}, we modify the quadratic coefficient of the local potential $f_{loc}(\phi)$ such that the local driving force for the transition varies across the sample. We assume that the local transition temperatures follow a Gaussian (Normal) distribution centered around the bulk transition temperature $T_0$:
\begin{equation}
P(T_c) = \frac{1}{\sigma \sqrt{2\pi}} \exp\left( -\frac{(T_c - T_0)^2}{2\sigma^2} \right)
\end{equation}
Here $T_0$ is the mean transition temperature of the pure stoichiometric crystal, and$\sigma$ represents the disorder strength (standard deviation), which controls the width of the phase coexistence region. A larger $\sigma$ leads to a more smeared transition and a wider hysteresis loop. The disorder potential $V_{dis}(\mathbf{r})$ is then defined by the local fluctuation in the $T_c$ field:
\begin{equation}
V_{dis}(\mathbf{r}) = \alpha \cdot \delta T_c(\mathbf{r})
\end{equation}
Here $\delta T_c(\mathbf{r}) = T_c(\mathbf{r}) - T_0$. 

In our numerical implementation, we can define a correlation length $\xi_{dis}$ for this disorder. If $\xi_{dis}$ is small (point defects), the landscape is jagged; if $\xi_{dis}$ is large (grain boundaries or strain domains), the landscape consists of "smooth hills and valleys" of stability. With the inclusion of $V_{dis}(\mathbf{r})$, the free energy functional now possesses multiple local minima distributed spatially. In hot spots ($T_c > T$), the metallic phase ($\phi = 1$) is locally stable. In cold spots ($T_c < T$), the insulating phase ($\phi = 0$) is locally stable. This effectively maps the Mott transition onto a Random Field Ising-like problem, where the random field is the local fluctuation in the chemical potential or $T_c$. This is the primary reason why $V_2O_3$ does not transform all at once; different regions flip at different temperatures, leading to the observed nanotextured phase coexistence.

In $V_2O_3$, the first-order metal-insulator transition is not purely electronic; it is a magnetostructural transition where the lattice symmetry shifts from rhombohedral (metallic) to monoclinic (insulating). This structural change introduces a significant lattice mismatch, creating long-range elastic strain that acts as a clamping force on the phase boundaries. To incorporate long-Range Elastic Interaction ($f_{el}$), we extend the Ginzburg-Landau functional to include the elastic energy and its coupling to the order parameter $\phi$. We need to take account for the fact that the structural distortion $\epsilon_{ij}$ is not independent; it must satisfy the Saint-Venant compatibility conditions to ensure the integrity of the crystal lattice. The elastic contribution to the free energy can be written as:
\begin{equation}
f_{el} = \frac{1}{2} C_{ijkl} \epsilon_{ij} \epsilon_{kl} - \eta \phi^2 \text{Tr}(\epsilon_{ij})
\end{equation}
Here, $C_{ijkl}$ is the elastic stiffness tensor, $\epsilon_{ij}$ is the strain tensor, and $\eta$ is the magneto-elastic coupling constant. This $\eta$ term couples the square of the order parameter (representing the phase fraction) to the local volume expansion or shear. 

By minimizing the elastic energy with respect to the displacement fields (and assuming mechanical equilibrium, $\nabla_j \sigma_{ij} = 0$), the strain can be integrated out. This transforms the local coupling into an effective long-range interaction between different regions of the order parameter:
\begin{equation}
F_{elastic} = \int \int d^3r d^3r' \Delta\phi(\mathbf{r}) K(\mathbf{r} - \mathbf{r'}) \Delta\phi(\mathbf{r'})
\end{equation}

Here $K(\mathbf{r} - \mathbf{r'})$ is an anisotropic kernel that falls off with distance (typically $1/r^3$). This represents the clamping effect. When a metallic grain ($V_{rhomb}$) tries to grow within an insulating matrix ($V_{mono}$), it exerts a long-range stress field that increases the energy cost for neighboring regions to transform. The competition between the Gaussian Disorder Landscape (which favors local, random transitions) and the Long-Range Elastic Interaction (which favors coordinated, long-range structural alignment) is the physical origin of the complex hysteresis in $V_2O_3$. The elastic interaction stiffens the energy landscape, making it harder for the system to reach global equilibrium. Instead of a single interface sweeping through the sample, the system breaks into a mosaic of self-accommodating strain domains (nanotexturing). As temperature decreases, the energy required to overcome these self-generated elastic barriers exceeds the available thermal energy $k_B T$, leading to the possibility of Kinetic Arrest of the transition.

We now move from the static energy landscape to the spatio-temporal dynamics. In $V_2O_3$, the arrest is fundamentally a kinetic phenomenon. The system possesses the thermodynamic driving force to transform, but the rate of transformation drops to zero. The evolution of the order parameter $\phi(\mathbf{r}, t)$ toward the minimum of the free energy functional $F[\phi]$ is governed by the TDGL equation :
\begin{equation}
\frac{\partial \phi(\mathbf{r}, t)}{\partial t} = -\Gamma(T, \mathbf{E}) \frac{\delta F}{\delta \phi(\mathbf{r}, t)} + \zeta(\mathbf{r}, t)
\end{equation}

In a standard phase transition, $\Gamma$ is often assumed to be a constant representing the material's inherent response rate. However, in a system exhibiting the signature of Kinetic Arrest, $\Gamma$ becomes a local, state-dependent variable that captures the viscosity of the energy landscape. We model the kinetic coefficient using an Arrhenius-like form to represent the thermally activated movement of the phase boundaries:
\begin{equation}
\Gamma(T, \mathbf{E}) = \Gamma_0 \exp\left( -\frac{U_{eff}(\phi, \mathbf{E})}{k_B T} \right)
\end{equation}
The effective barrier $U_{eff}$ is a composite of three critical energies:(i) $U_{int}$ (intrinsic): the electronic energy barrier associated with the Mott gap and the structural reconfiguration (rhombohedral-monoclinic); (ii) $U_{el}$ (elastic clamping): as we developed in the previous section, the long-range strain fields created by growing metallic droplets act as a pinning force. $U_{el}$ increases as the phase fraction $\phi$ changes, creating a self-limiting growth mechanism; (iii) $W_{ext}$ (external field): the external electric or magnetic field $\mathbf{E}$ provides a bias that can either lower the barrier (promoting the transition) or lock the current state (arresting the transition).

The transition gets arrested when the thermal energy $k_B T$ is no longer sufficient to overcome the total barrier. At high $T$, $\Gamma$ is large, and the system rapidly moves toward the global minimum of $F[\phi]$. The transition follows standard thermodynamics. At low $T$ ($T \rightarrow T_K$), as the temperature drops toward the kinetic arrest temperature ($T_K$), the exponential term $\exp(-U/k_B T)$ collapses toward zero. Even though the free energy functional $F[\phi]$ still shows that the insulating phase is thermodynamically preferred, the velocity of the interface $v \propto \Gamma \frac{\delta F}{\delta \phi}$ vanishes.

The field-Modified aspect is what explains the memristive behavior observed experimentally in V$_2$O$_3$. By applying an external field $\mathbf{E}$, we effectively modify the landscape:
\begin{equation}
U_{eff} \rightarrow U_{eff} - \mathbf{p} \cdot \mathbf{E}
\end{equation}

Here $\mathbf{p}$ is the local dipole or strain-coupling moment. This allows for a system that was kinetically arrested (frozen) at a low temperature can be de-arrested or devitrified by an external field that lowers the barrier enough for the kinetics to restart. This field-induced de-arrest is the physical basis for the resistive switching in $V_2O_3$ memristors.

In the context of a first-order phase transition (FOPT) influenced by quenched disorder, the noise term $\zeta(\mathbf{r}, t)$ is the engine of nucleation. Without it, a system trapped in a local minimum (a metastable state) would remain there forever, even if a lower energy state exists. For the model to be thermodynamically consistent, the noise must be Gaussian white noise that satisfies the Fluctuation-Dissipation Theorem (FDT). This ensures that the noise strength is proportional to the damping (the kinetic coefficient $\Gamma$):
\begin{equation}
\langle \zeta(\mathbf{r}, t) \zeta(\mathbf{r'}, t') \rangle = 2 \Gamma(T, \mathbf{E}) k_B T \delta(\mathbf{r} - \mathbf{r'}) \delta(t - t')
\end{equation}

The term $2 \Gamma k_B T$ signifies that the same microscopic processes responsible for slowing down the phase boundary (dissipation) are also responsible for the random thermal kicks (fluctuations). In the $V_2O_3$ system, the noise facilitates the transition across the energy barriers $U_{eff}$. Above $T_K$ (ergodic regime), thermal fluctuations are large enough to overcome the barriers. The system can hop out of metastable local minima, allowing the metallic or insulating phases to nucleate and grow toward the global equilibrium state. As temperature $T$ decreases, two things happen simultaneously. The noise amplitude $\sqrt{2\Gamma k_B T}$ shrinks. The mobility $\Gamma$ itself collapses exponentially (due to the $U_{eff}$ barrier). Below $T_K$ (Kinetic Arrest), the kicks provided by $\zeta$ become so infrequent and weak that the system can no longer escape its local environment. The system's state becomes frozen or arrested.

Because the noise $\zeta(\mathbf{r}, t)$ acts on a landscape already roughened by Gaussian Disorder ($V_{dis}$), the nucleation is not uniform. In regions where the local $T_c$ is favorable (the hot spots), even small noise fluctuations can trigger the transition. In clamped regions (high elastic strain), even large fluctuations fail to move the phase boundary. Incorporating $\zeta$ allows on to argue that the metastability in $V_2O_3$ is a result of a competition between time scales: (i) The Internal Relaxation Time ($\tau_{rel} \sim 1/\Gamma$); (ii) The Experimental Observation Time ($t_{exp}$). When $\tau_{rel}$ exceeds $t_{exp}$ due to the collapse of $\Gamma$ and the weakening of $\zeta$, the system enters the kinetically arrested state.

The TDGL equation provides the local phase evolution $\phi(\mathbf{r}, t)$. We niow define the global transformed fraction $X(t)$ as the spatial average of the order parameter:
\begin{equation}
X(t) = \frac{1}{V} \int_V \phi(\mathbf{r}, t) \, d^3r
\end{equation}

In the standard KJMA model (see \cite{kiana} and references within), the fraction of transformed volume is given by:
\begin{equation}
X(t) = 1 - \exp(-Kt^n)
\end{equation}

By running TDGL simulations on a grid with the Gaussian disorder landscape ($V_{dis}$) and long-range elastic interaction ($f_{el}$), we can numerically determine the constants $K$ and $n$.TThe rate constant ($K$) is directly proportional to the kinetic coefficient $\Gamma(T, \mathbf{E})$. In the arrested regime of $V_2O_3$, $K$ drops exponentially, signifying that the velocity of the phase transformation has stalled. The Avrami exponent ($n$) reflects the geometry and dimensionality of the growth. In an ideal, clean system, $n$ might be 3 or 4 (3D spherical growth).  In the presence of quenched disorder and elastic clamping, the growth is hindered. The TDGL simulation will show that the metallic domains become jagged and anisotropic, leading to an effective $n < 1$. This matches the experimental observation of slow or "stretching" kinetics in $V_2O_3$.

The core of KJMA model is the concept of extended volume. It is the volume the new phase would occupy if grains could overlap and nucleation was random. In our modified TDGL framework, the nucleation sites are not random; they are dictated by the hot spots  in the Gaussian $T_c$ distribution. Growth is not constant; it is slowed by the $U_{eff}$ barriers (Kinetic Arrest). The bridge is established by showing that the effective activation energy $U_{eff}$ from our TDGL formalism is what dictates the temperature-dependence of the KJMA rate constant $K(T)$.

\begin{figure}[t]
\centering
\includegraphics[scale=0.4]{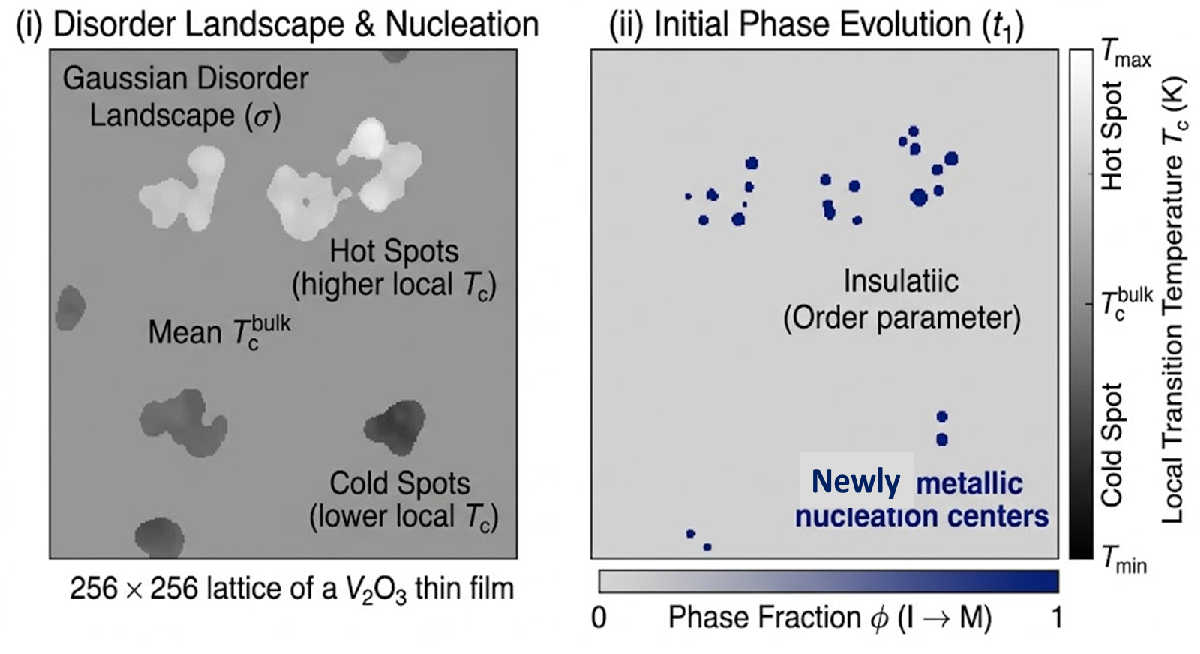}
\caption{Quenched disorder and localized nucleation": (i) disorder landscape and nucleation; (ii) initial phase evolution. }
\end{figure}

To visualize the initial field-induced insulator-to-metal transition (IMT), we solved the discretized TDGL equation on a $256 \times 256$ lattice, simulating a characteristic cross-section of an epitaxial $V_2O_3$ film. Initialized in the insulating phase ($\phi \approx 0$) at $T < T_c^{bulk}$, the system was subjected to a sub-threshold field ($\mathbf{E}$) to isolate the role of quenched disorder.Figure 1 illustrates the deterministic link between microscopic material landscape and the onset of the transition. Panel (i) maps the Gaussian disorder landscape ($\sigma$), where intensity gradients represent variations in the local transition temperature $T_c(\mathbf{r})$; lighter regions denote "hot spots" with locally higher $T_c$, making them thermodynamically favorable for nucleation. Panel (ii) provides a spatial map of the order parameter $\phi(\mathbf{r})$ at time $t_1$. A direct 1:1 comparison confirms that metallic nucleation centers (dark blue) emerge exclusively at these light-colored hot spots. This visual correlation demonstrates that the unique realization of quenched disorder in a sample dictates the initial droplet geometry, effectively smearing the sharp bulk transition into a continuous, non-equilibrium avalanche process typical of strained Mott insulators.

\begin{figure}[t]
\centering
\includegraphics[scale=0.4]{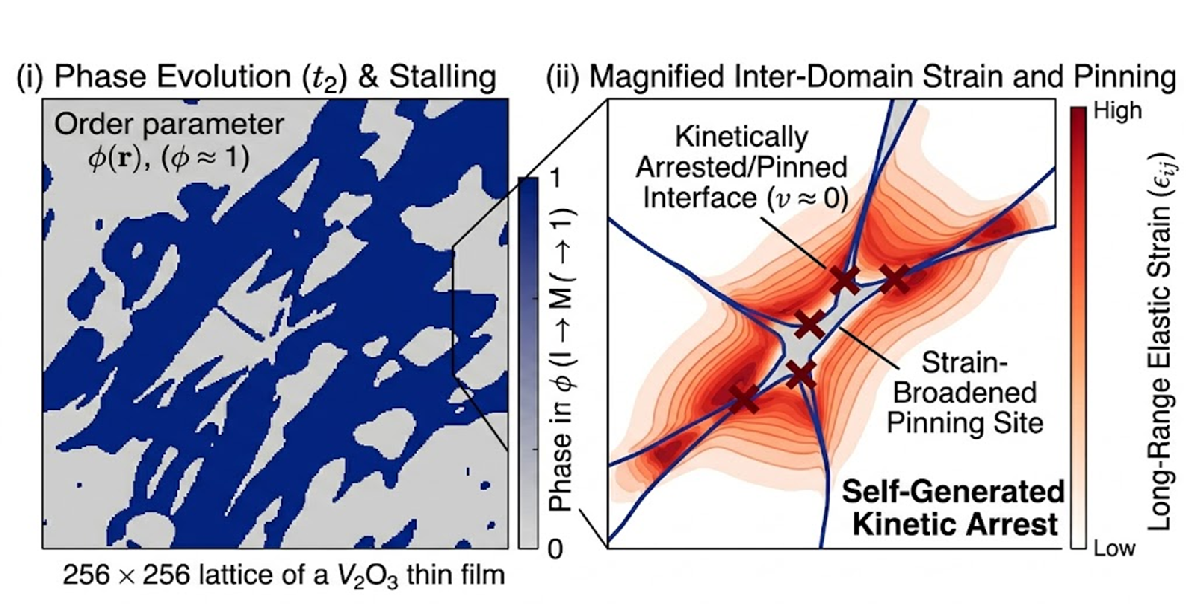}
\caption{Magnified inter-domain strain and pinning: (i) phase evolution and stalling; (ii) magnified inter-domain strain and pinning. }
\end{figure}

Following nucleation, the morphology and kinetics of $V_2O_3$ are governed by long-range elastic interactions. The mismatch between the rhombohedral metallic phase and the monoclinic insulating matrix generates anisotropic strain fields ($\epsilon_{ij}$), causing expanding domains to deviate from isotropic growth. Figure 2 captures the system at a later time $t_2$, illustrating the emergence of nanotexturing and growth stalling. As shown in Panel (i), the matured metallic domains (dark blue) exhibit jagged, plate-like boundaries, reflecting the minimization of total elastic energy along preferred crystallographic directions. Unlike ideal systems, the transition here is visibly arrested at a fractional completion $X < 1$. Panel (ii) demonstrates the physical mechanism: as domains approach one another, their overlapping strain fields (red/orange contours) create an insurmountable kinetic barrier, $U_{eff}$. This interface pinning (red 'X') reduces the boundary velocity $v$ toward zero, trapping the system in a non-equilibrium mosaic of persistent phase coexistence. This real-space visualization of KA provides the physical basis for the incomplete resistivity switching observed experimentally, proving that arrest is a self-generated, strain-induced bottleneck rather than a mere consequence of static defects.

\begin{figure}[t]
\centering
\includegraphics[scale=0.4]{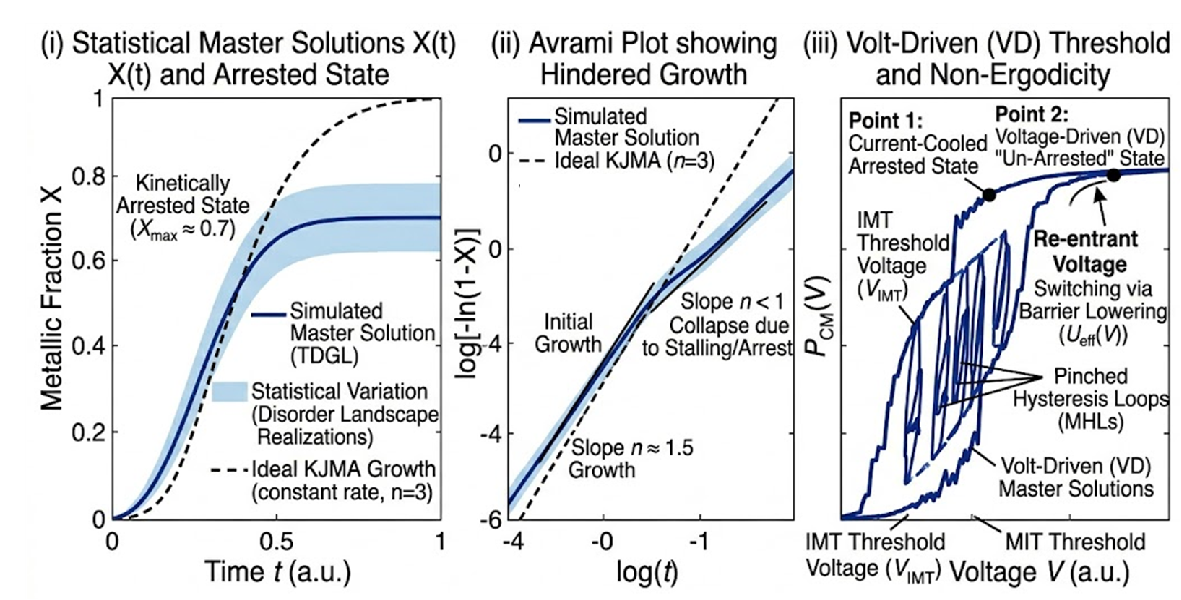}
\caption{Statistical master-solutions and non-linear kinetics: (i) Statistical master-solutions X(t) and arrested state; (ii) Avrami plot showing hindered growth; (iii) voltage driven threshold and non-ergodicity. }
\end{figure}

Figure 3 synthesizes the transition kinetics by comparing our generalized TDGL framework with experimental signatures. Panel (i) plots the global metallic fraction $X(t)$, contrasting ideal KJMA growth ($n=3$, dashed black), which saturates at $X=1$, with our simulated master solution (solid blue). The latter exhibits a sigmoidal profile that stalls at $X \approx 0.7$, providing statistical confirmation of the persistent phase coexistence and kinetic arrest (KA) visualized in Fig. 2. The shaded band represents the statistical broadening across multiple disorder realizations, reflecting the stochastic nature of nucleation at random "hot spots."Panel (ii) presents the Avrami analysis, $\log[-\ln(1-X)]$ vs. $\log(t)$. While ideal growth maintains a constant slope ($n=3$), our model shows a critical collapse in the slope as the system approaches the stalled state, with the Avrami exponent $n$ dropping below unity. This indicates interface pinning due to cumulative strain. Furthermore, the Volt-Driven (VD) Master Solutions demonstrate that an arrested state can be "de-arrested" by increasing the external field, which lowers the local barriers $U_{eff}$. This field-dependent de-arresting of individual domains generates the stable, multi-level resistance states and minor hysteresis loops (MHLs) observed in experiments. The threshold voltage $V_{th}$ is found to scale exponentially with the effective activation barriers $U_{eff}$ derived from our theory, establishing a direct link between microscopic strain-clamping and macroscopic memristive functionality.

In summary, we have developed a generalized Time-Dependent Ginzburg-Landau (TDGL) framework that provides a unified microscopic description of field-induced first-order phase transitions (FOPT) in the presence of quenched disorder and long-range elastic interactions. By applying this model to the canonical Mott insulator $V_2O_3$, we have demonstrated that the observed slow kinetics and persistent phase coexistence are not mere experimental artifacts, but fundamental consequences of self-generated kinetic arrest (KA). Our numerical results provide three critical insights into the physics of correlated materials. We show that the smearing of the Mott transition is governed by a Gaussian disorder landscape, where nucleation is deterministically tied to local hot spots. We identify long-range strain as the primary mechanism for interface pinning. The resulting collapse of the Avrami exponent ($n < 1$) serves as a universal signature of hindered growth in epitaxial films. By formalizing the field-dependence of the kinetic coefficient $\Gamma(T, \mathbf{E})$, we show that memristive switching is an ergodicity-breaking process. The applied field lowers the effective activation barriers $U_{eff}$, allowing the system to de-arrest and navigate a manifold of metastable, non-equilibrium states. Beyond $V_2O_3$, this framework is readily applicable to a broad class of materials exhibiting glassy first-order transitions, including manganites, ferroelectrics, and layered ferromagnets (such as the FeGeTe family). By bridging the gap between microscopic field theory and macroscopic KJMA statistics, our work provides a predictive tool for designing the next generation of neuromorphic and energy-storage devices. The ability to precisely tune the arrested state via strain engineering and external bias opens a new frontier in the control of quantum materials far from equilibrium. It may be noted that the important role of strain in the first order magneto-structural phasd transition has been flagged in earlier experimental studies \cite{kalch2020,sathe2014}

\end{document}